\documentclass[twocolumn,showpacs,preprintnumbers,amsmath,amssymb,aps,prb,floatfix,groupedaddress]{revtex4}
\usepackage{graphicx} 
\usepackage{dcolumn} 
\usepackage{bm} 
\usepackage{epsfig}
\usepackage{color}
\usepackage{multirow}

\usepackage{amsmath}

\begin{document}

\title{Dirac Point Degenerate with Massive Bands \\at a Topological Quantum Critical Point}
\author{J. C. Smith}
\affiliation{Department of Physics, University of California Davis}
\author{S. Banerjee}
\affiliation{Department of Physics, University of California Davis}
\author{V. Pardo}
\affiliation{Department of Physics, University of California Davis}
\author{ and W. E. Pickett}
\affiliation{Department of Physics, University of California Davis}

\begin{abstract}
The quasi-linear bands in the topologically trivial skutterudite insulator CoSb$_3$
are studied under adiabatic, symmetry-conserving displacement of the Sb sublattice.
In this cubic, time-reversal and inversion symmetric system, a
transition from trivial insulator to topological point Fermi surface system
occurs through a critical point in which massless (Dirac) bands are {\it degenerate}
with massive bands.  Spin-orbit coupling does not alter the character of the
transition.  The mineral skutterudite (CoSb$_3$) is very near the critical point
in its natural state.
\end{abstract}
\date{\today}
\maketitle

\vskip2pc


The topological properties of crystalline matter have become a central feature 
in characterizing the electronic structure of small gap, primarily binary, 
semiconductors.\cite{NiuReview,KaneReview,Moore,Zhang} 
Skutterudite compounds, many of which have small gaps, have received a great deal of interest
in the past two decades.  Most recently emphasis has been on
the ``filled'' version in which atoms are incorporated in the large holes in the
original skutterudite (CoSb$_3$) structure, which can become
unusual heavy fermion correlated metals and even superconductors.\cite{maple}
The earlier interest was in their transport properties.\cite{caillat}  As small
gap semiconductors, many of them were of potential interest in
solid state devices, and application as thermoelectric materials\cite{tritt,sales}
was a strong interest.  

A study of the electronic structure\cite{djswep} uncovered a very peculiar
feature of some of them: there are linear valence and conduction
bands that extended from well out in the Brillouin
zone, changing to quadratic only very near the zone center $k$=0.  
This quasi-linear dispersion produces
peculiar consequences: the density
of states behaves as $\varepsilon^2$ near the band edge rather than
the usual three dimensional (3D) form $\sqrt{\varepsilon}$; the
carrier density scales differently with Fermi energy $\varepsilon_F$;
the inverse mass tensor $\nabla \nabla \varepsilon_k$ is entirely
off-diagonal corresponding to an ``infinite'' transport mass; the
cyclotron mass is different from usual 3D behavior, etc.  All of
this was unique and was potentially very useful in applications, 
but theoretical excitement was tempered
because the quasilinear dispersion, which was clearest in IrSb$_3$,
finally became quadratic {\it very near} $k$=0, just as textbooks
claim must be the case.

Since then, graphene has been isolated and its ``Dirac point'' with
linear dispersion has been studied comprehensively.\cite{graph1,graph2}  The Dirac point
of graphene however occurs at a zone corner point where symmetry is
much lower than at the zone center, and its occurrence
does not violate textbook conventional wisdom.  Here we show
that in the skutterudite system 
small adjustments in the structure produce a critical point at which strictly linear bands
extrude from $|\vec k|$=0.  This does not violate any real principle,
however it does violate the commonly used expansions.
The linear behavior reflects non-analytic behavior in the
$\vec k \rightarrow 0$ limit, resulting from an accidental (but
tunable) degeneracy.  In this paper we illustrate how to tune to this
critical point, provide a simple model that
reproduces the behavior, and demonstrate that the transition corresponds
to a topological transition as well.


\noindent
\begin{figure}[!htb]
\begin{center}
\includegraphics[width=0.32\textwidth,angle=0]{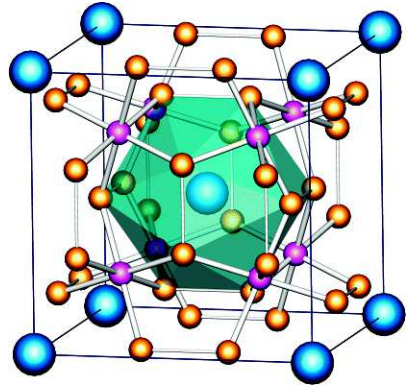}
\end{center}
\caption{(Color online)
Crystal structure of skutterudite CoSb$_3$,
space group $Im\bar{3}$ (\#204).  The experimental lattice constant is
$a$=9.0385 $\AA$ and the internal position coordinates are $u$=0.335, $v$=
0.1575.  The Co site (small pink sphere) is octahedrally coordinated to
Sb atoms (small yellow spheres), each of which connects two octahedra.  The
large (blue) sphere denotes a large open site which is unoccupied in CoSb$_3$;
the surrounding solid (center of figure) gives an idea of the volume and shape of the empty region. 
\label{Structure}
}\end{figure}

The skutterudite structure, pictured in Fig.~\ref{Structure}, in the
space group $Im{\bar 3}$ (\#204), has a simple cubic Bravais
lattice, and is comprised a bcc repetition of four formula units (f.u.) when
expressed at $TPn_4$.  The pnictide ($Pn$) atoms form bonded units (nearly square but 
commonly designated as rings) which are not required
by local environment or overall symmetry to be truly square; therefore
they are not although very nearly so.  The three Pn$_4$ squares in the primitive
cell as oriented perpendicular to the coordinates axes.  Four transition metal ($T$)
atoms (Co, Ir, ...) lie in four of the subcubes of the large cube of lattice
constant $a^3$; the other four subcubes (octants) are empty.  The structure
is symmorphic, with 24 point group operations; the one that is missing is
reflection in (110) planes.  This space group leads to some interesting band
behavior but is not relevant to the behavior we discuss in this paper.  The related 
filled skutterudites $XT_4$Pn$_{12}$ have an atom $X$ incorporated into the large
$2a$ site of $3\bar m$ symmetry. 

A relevant structural feature is that skutterudite is related to the perovskite
structure $\Box T$Pn$_3$ ($\Box$ denotes an empty A site). Beginning from 
perovskite, a rotation of the octahedra keeping the Pn atoms along the cube
faces results in the formation of the (nearly square) Pn$_4$ rings, and the 
Pn octahedra become distorted and less identifiable as a structural feature. The
transformation is, in terms of the internal coordinates $u$ and $v$,
\begin{eqnarray}
u' = \frac{1}{2} + s (u - \frac{1}{2});
v' = \frac{1}{2} + s (v - \frac{1}{2}).
\end{eqnarray}
The transformation path, from perovskite for $s$=0 to the observed structure
for $s$=1, is pictured in Fig. 1 of Ref. \onlinecite{Llunell}.
Below we make
use of this transformation to understand the opening of the (pseudo)gap between
occupied and unoccupied states.  

{\it Evolution through a critical point.}
The electronic structure of skutterudites has been of keen interest since the 
quasi-linear bands (QLB) near the zone center were uncovered by Singh and Pickett.\cite{djswep}
The skutterudites that are isovalent with CoSb$_3$ are very narrow gap semiconductors
(or possibly very small negative gap semimetals, or point Fermi surface zero-gap
materials, {\it viz.} IrSb$_3$~[\onlinecite{djswep}]).  In following the band structure
along the perovskite-to-skutterudite structural path given above, it is found that the
gap only opens up near the end  of the transformation ($s \sim 0.90-0.95$), where the
Sb$_4$ rings approach their equilibrium size and the empty $2a$ site is fully 
developed into a large interstice.  Only near $s \sim 1$ does the quasilinear band 
emerge from the dense spaghetti of occupied valence Sb $4p$ and Co $3d$ bands.
Analysis of the band character and projected density of states (DOS) indicates no
Co $3d$ character and very little Sb $4p$ character in the quasilinear bands, which
therefore arise from Bloch states (one on either side of the gap) associated with 
electrons in the large empty $2a$ site.  Analogous bands have been observed in other
open lattices, for example in the Cs crown ether molecular solid.\cite{crownether} 

\noindent
\begin{figure}[!htb]
\begin{center}
\rotatebox{-90}{\includegraphics[width=0.30\textwidth,angle=0]{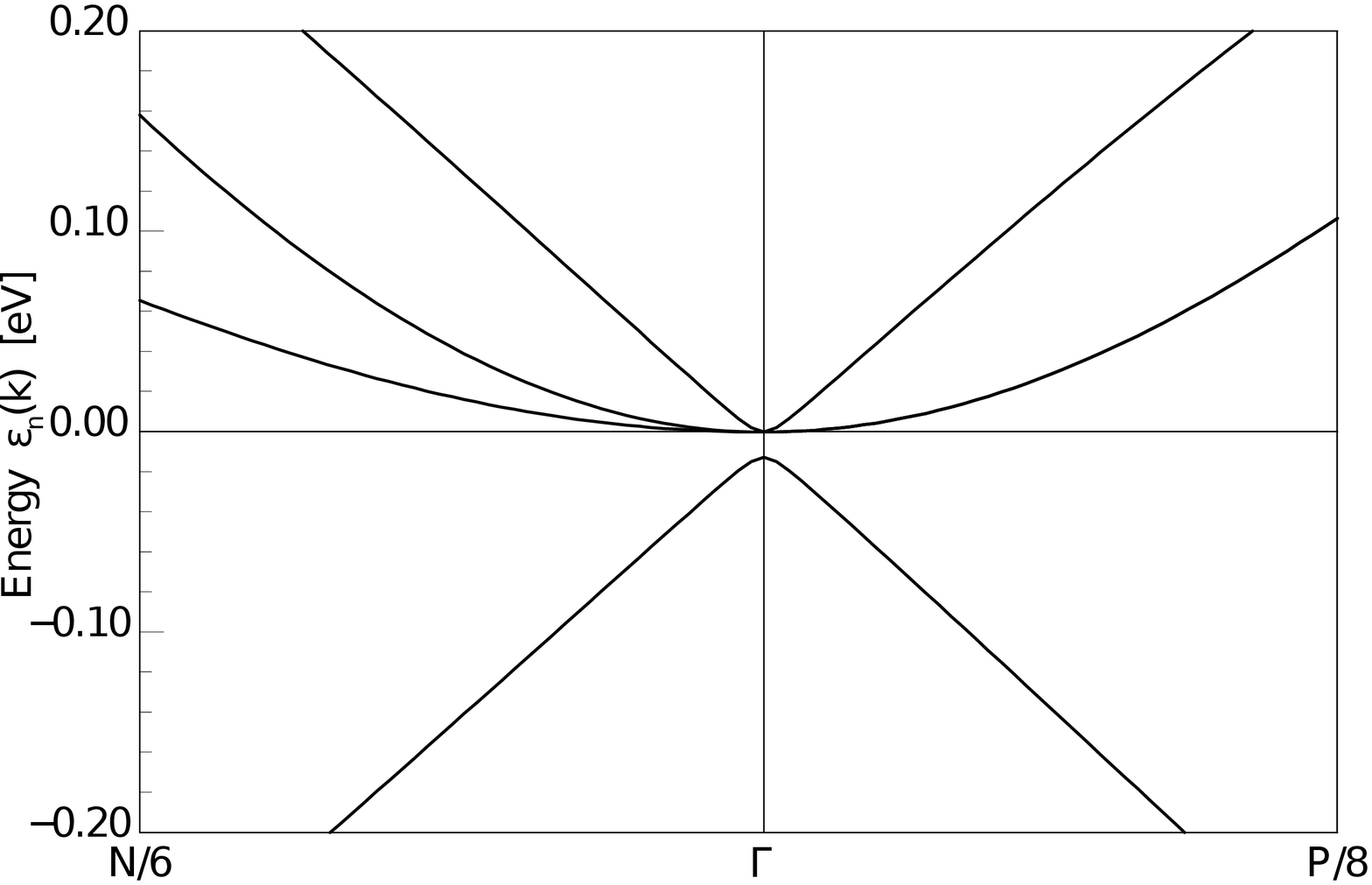}}
\rotatebox{-90}{\includegraphics[width=0.30\textwidth,angle=0]{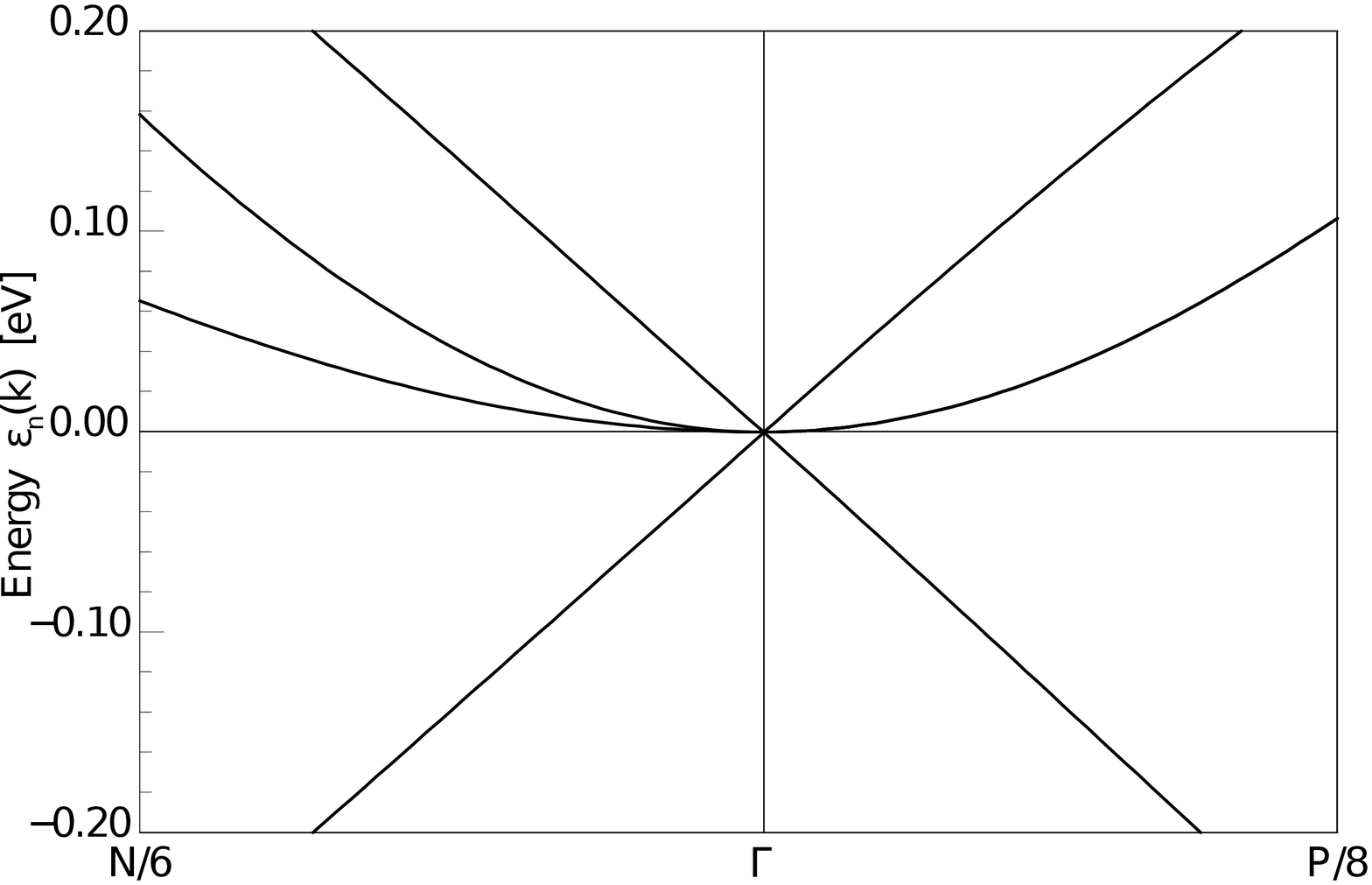}}
\rotatebox{-90}{\includegraphics[width=0.30\textwidth,angle=0]{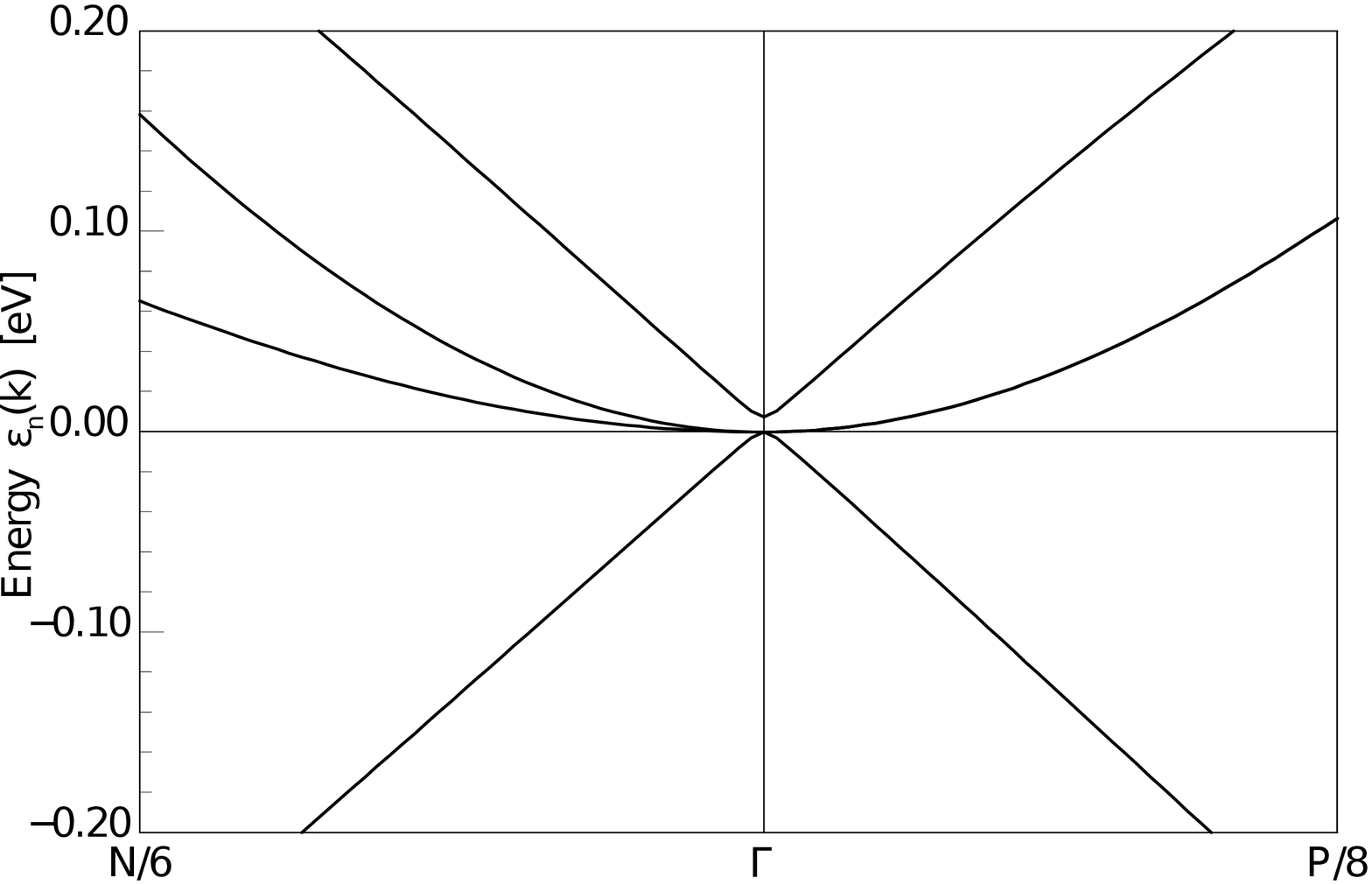}}
\end{center}
\caption{
Bands near $k$=0 in slightly distorted skutterudite CoSb$_3$, showing the
band crossing (as the valence band rises due to the shift of the Sb position) 
through the critical point of quadruple degeneracy of a
Dirac pair and a conventional band pair. Top: before transition. Middle: at
the critical point. Bottom: just after the transition.
\label{BandsNoSOC}
}\end{figure}

To illustrate the progression of the band structure through a critical point at which 
a Dirac point (with Dirac hypercone) appears, we provide in Fig. \ref{BandsNoSOC}
the behavior of the bands for $s =$ 1.020, 1.023, 1.025, corresponding to just before,
precisely at, and just beyond gap closing.  At zero gap, the QLBs become precisely linear
(Dirac) bands.  Because one of them is degenerate (by crystal symmetry) with
two other bands in the 3-fold set, this Dirac point is degenerate with two conventional
(massive) conduction bands.  Beyond the critical point $s_{cr}=1.023$, the singlet
lies above the triplet, and the Fermi level lies at a symmetry-determined, point
FS energy comprised of one hole and two electron bands.  While beyond the critical
point these bands are all
``massive'' in the rigorous sense, immediately beyond the transition the masses of both 
quasilinear (valence and conduction) bands {\it arise continuously from zero mass}
to the linear behavior that extends as far as the bands can be followed 
before they helplessly mix with and disappear into the background spaghetti.

\noindent
\begin{figure}[!htb]
\begin{center}
\rotatebox{-90}{\includegraphics[width=0.30\textwidth,angle=0]{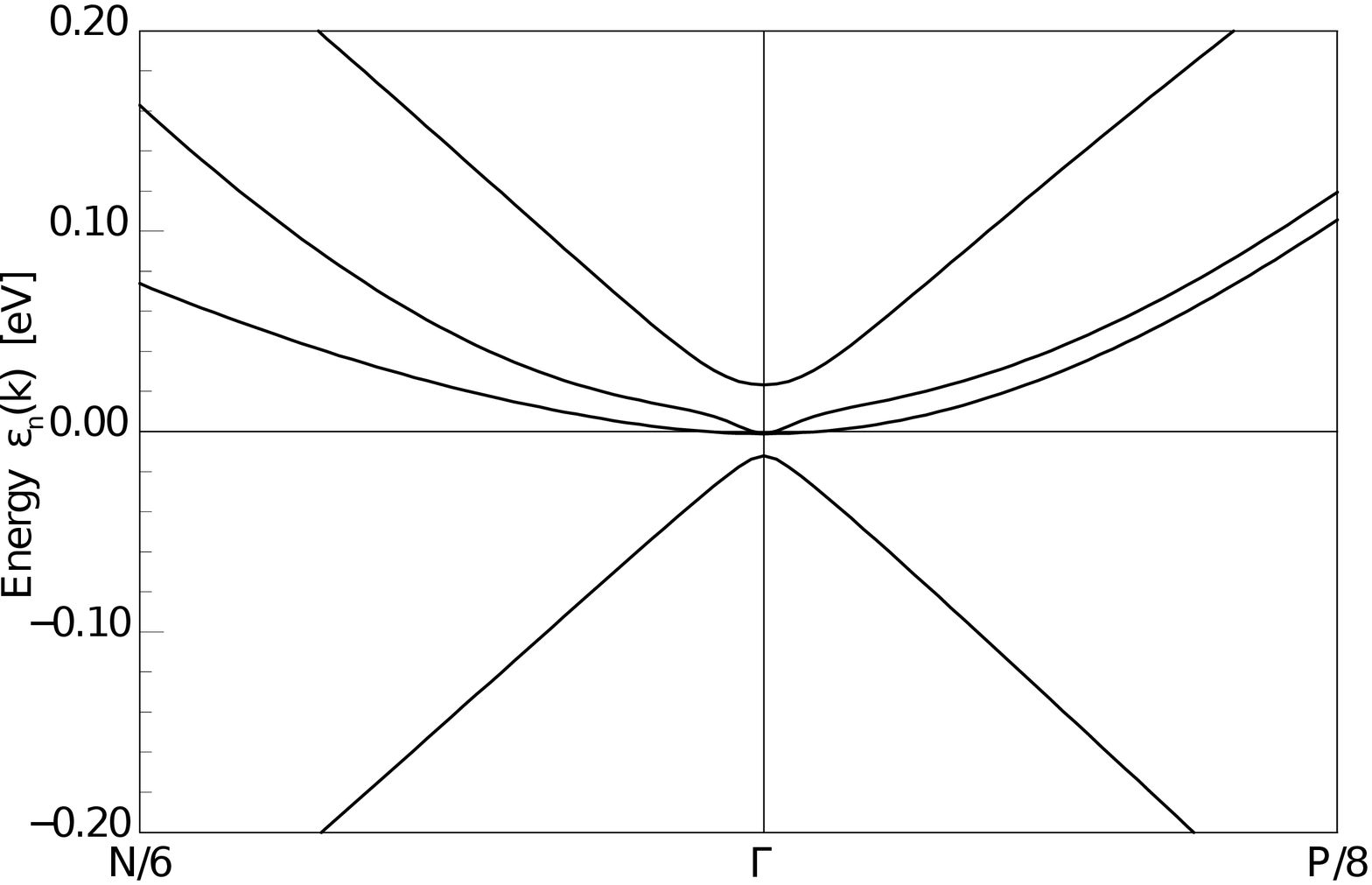}}
\rotatebox{-90}{\includegraphics[width=0.30\textwidth,angle=0]{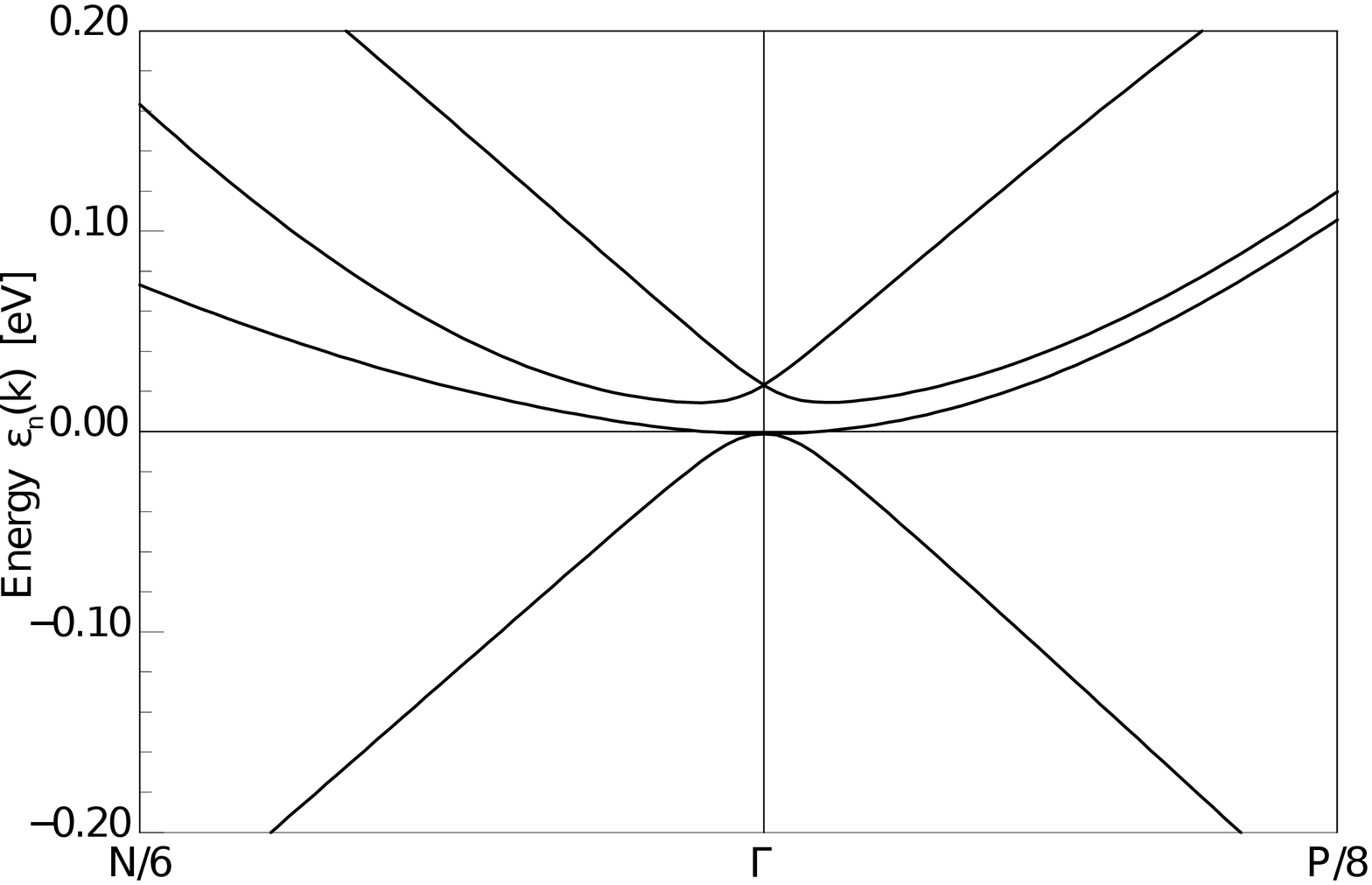}}
\rotatebox{-90}{\includegraphics[width=0.30\textwidth,angle=0]{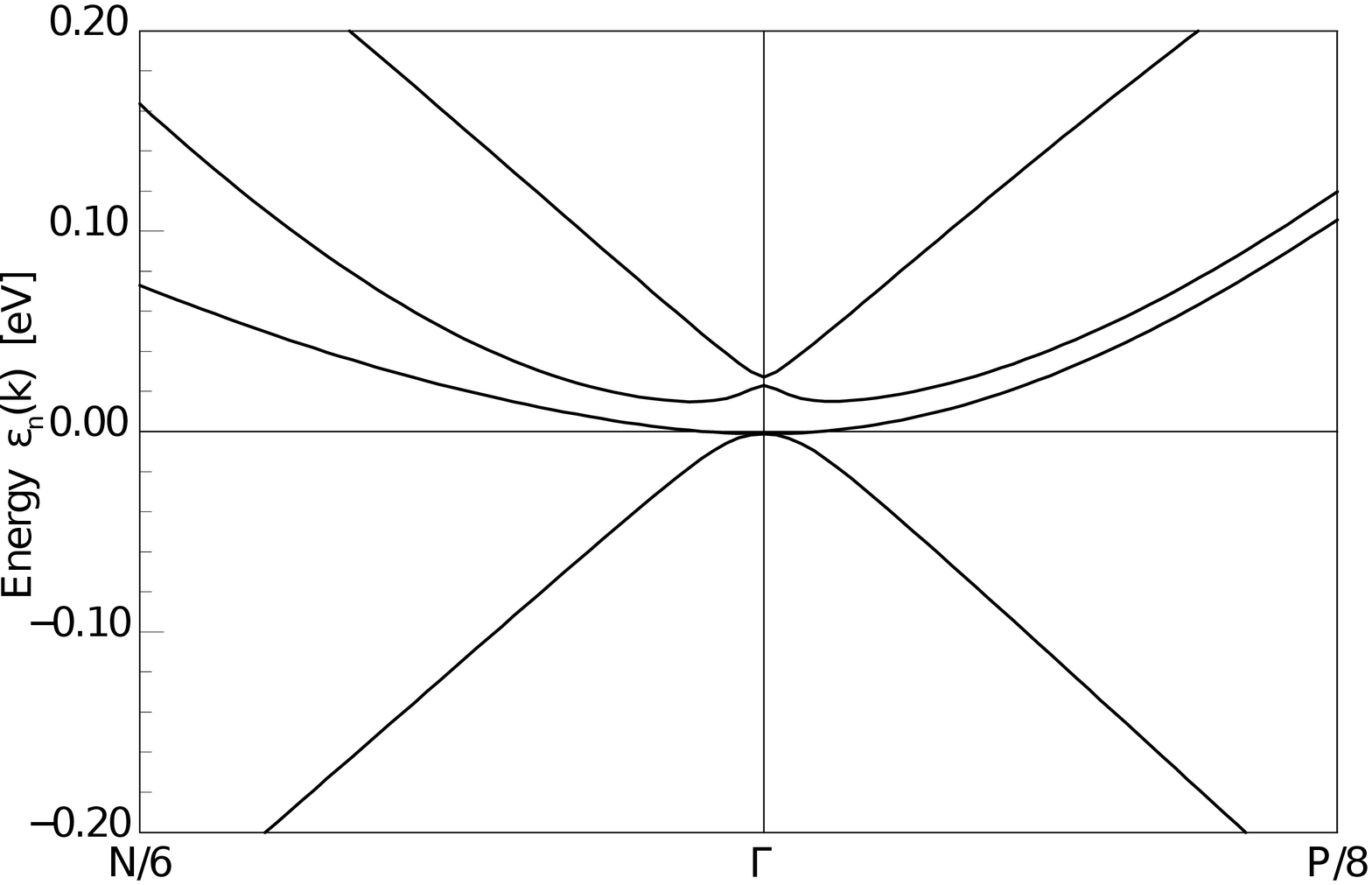}}
\end{center}
\caption{
Bands as in Fig. \ref{BandsNoSOC}, with $s$=1.010, 1.019, 1.020 and with spin-orbit coupling included.
Although the threefold ``p'' band degeneracy is split by SOC, the Dirac
bands and hypercone survive, though the lower (hole) band mixes with one of the massive
bands very close to $k$=0.
\label{BandsWithSOC}
}\end{figure}

At first sight, the basic underlying feature seems to be provided by two 
states at $\pm \varepsilon_{\circ}$ on some scale,
which will become degenerate ($\varepsilon_{\circ}=0$) at the critical point. Although
the two-band\cite{Sofo} Kane model has been used 
to represent the bands of CoSb$_3$, it fails
to give the linear dispersion at {\it arbitrarily small $k$} as the gap vanishes, so
some other picture must be constructed.  
While the bands are required
to have only cubic -- not spherical --symmetry, the bands for CoSb$_3$ are in fact isotropic,
that is, the velocity is indistinguishable in all three high symmetry directions.
The simplest viewpoint is that two bands are linearly coupled ($h_{ij} \propto v|\vec k|$
for $i\neq j$)
at small $|\vec k| \equiv k$, in which case the eigenvalues are
\begin{eqnarray}
\varepsilon_k = \pm \sqrt{\varepsilon_{\circ}^2 + (v|\vec k|)^2} 
             \rightarrow& \pm v|\vec k|,
\end{eqnarray}
giving the desired two linear bands upon degeneracy ($\varepsilon_{\circ}\rightarrow 0$).

So how does one obtain the desired coupling? The easiest way to get linear coupling at small $k$, 
in a tight-binding picture,
is from a coupling such as $t$(sin$k_x a$ + sin$k_y a$ + sin$k_z a$) on the off-diagonal.  
However, expanding this
coupling for small $k$ ($k_x + k_y + k_z$) does not give isotropic coupling.  What could give isotropic
coupling?  

The skutterudite structure, which has bcc translational symmetry with coupled Sb$_4$ ring
5$p$ orbitals and large empty holes in
the lattice that may harbor an $s$-like orbital in its well, can be modeled with 
a $p$ triplet coupled to $s$-symmetry states in the open holes.  Working in a
picture where the $p$ triplet is diagonalized at $k$=0, 
the coupling of the $p_x$ function with the bcc-situated $s$ orbitals gives
a nearest-neighbor coupling of
\begin{eqnarray}
T_x \equiv T(k_x,k_y,k_z) = 8it~sin\frac{k_xa}{2}cos\frac{k_ya}{2}cos\frac{k_za}{2}
\end{eqnarray}
and symmetrically for coupling of $p_y$ and $p_z$ partners.  Then using on-site
energies $\varepsilon_s$ and $\varepsilon_p$, the tight-binding Hamiltonian is
\begin{eqnarray}
H = \left(\begin{array}{cccc}
  \varepsilon_s & T_x  & T_y  & T_z  \\
   T^*_x & \varepsilon_p  &  0  & 0  \\
   T^*_y &  0  & \varepsilon_p &  0  \\
   T^*_z &  0  & 0  & \varepsilon_p  \end{array}\right)
\label{ham}
\end{eqnarray}
with eigenvalues
\begin{eqnarray}
\varepsilon_j =\varepsilon_p; 
               \varepsilon_p; 
      \frac{\varepsilon_s + \varepsilon_p}{2} 
    \pm \sqrt{(\frac{\varepsilon_s - \varepsilon_p}{2})^2
   + |T|^2}, 
\end{eqnarray}
where $|T|^2 \equiv |T_x|^2 + |T_y|^2 + |T_y|^2$.  To first order in $k$ and
and at the critical point $\varepsilon_s \rightarrow \varepsilon_p$, this result
gives (1) 4-fold degenerate bands $\varepsilon=\varepsilon_p$ at $k$=0 (where $T$ vanishes), 
(2) two bands have isotropic 
linear dispersion $\varepsilon_p \pm vk$ with $v = 4ta$, 
(3) the other two bands are flat in Eq. \ref{ham},
but will acquire finite mass by the smaller $p-p$ hopping that has been neglected for simplicity.
For $|\varepsilon_p \sim
\varepsilon_s|$, three-fold degeneracy is preserved at $k$=0. 
This model faithfully reproduces the behavior
in CoSb$_3$ in Fig. \ref{BandsNoSOC} as the Sb rings are varied in size adiabatically.

A number of workers\cite{Fu1,murakami,roy} have pointed out that insulators in 
3D, as well as in 2D, can be
characterized by topological invariants, and Fu and Kane followed by demonstrating\cite{Fu2}
that when inversion symmetry is present (as in space group $Im\bar 3$), the Z$_2$
invariant can be obtained from the parities of the occupied states at the invariant momenta,
which in the $bcc$ structure consist of $\Gamma$, three H points [viz. (2,0,0)
$\frac{\pi}{a}$], and the four P points [viz. (1,1,1)$\frac{\pi}{a}$]. Here only
the $\Gamma$ point requires consideration, since reoccupation occurs only there.
The lower band in Fig. \ref{BandsNoSOC} has {\it odd} parity at $\Gamma$ while the 
triplet is {\it even}.  As the critical point is crossed, the product of the parities of the
occupied bands at $\Gamma$, and hence the Z$_2$ invariant, changes sign, 
the signal of a trivial to
topological transition. This change also reveals that the transition is associated with the
entanglement of the odd symmetry valence band with the even conduction band that has the
same symmetry away from $\Gamma$, and hence mixes with.  
The final state is actually gapless; it is a 
(point Fermi surface) zero-gap semiconductor, with the mass of the lowest band rising 
from zero and giving rise to extremely light mass carriers in the limit of low
hole doping.  The system could be rendered a true topological insulator by strain 
(lifting of the band degeneracy).

{\it Effect of Spin-Orbit Coupling.}
For topological states in crystals, spin-orbit coupling (SOC) has been a central issue.
In Fig. \ref{BandsWithSOC} the effect of intrinsic (relativistic) SOC is shown.  The triplet is split 
(by 40 meV) into a
lower energy doublet and higher energy singlet, as long as the gap exists.  At the critical
point $s_{cr}^{soc}=1.019$ (it is slightly shifted by SOC)
the (formerly) valence band singlet has crossed 
the two-fold level and become
degenerate with the conduction singlet, giving rise to a Dirac point involving the 
two upper bands which are now separated from the doublet.  
While the details have changed somewhat, the initial gapped state, the Dirac point
at critical distortion, and the zero-gap final state remain. SOC therefore produces no
qualitative change in the transition.

We have established that the trivial insulator to topological zero-gap semiconductor
occurs simultaneously with the appearance of a Dirac point at $k$=0, which is degenerate
with conventional (massive) bands at the critical point.  The appearance of the Dirac
point at $k$=0 is clarified, being due to the tuning of a degeneracy of site energies of the orbitals
that are involved.  A uniaxial strain, externally applied or resulting from epitaxial
growth on a substrate with some lattice match, will lift the remaining degeneracy and 
produce the topological insulating state that has so far attracted the main interest
in this area.  

It is worthwhile to note that this ``robust'' topological state is actually delicate
with respect to the Sb sublattice position: the transition occurs discontinuously at
$s = s_{cr}$ upon continuous, symmetry-preserving change of the Sb coordinate.  Such a situation
will allow probing into just which (bulk or surface) properties are associated with
the topological nature of the bulk electronic state.  Of course, there are many properties
that change discontinuously at an insulator-to-metal transition, so effects of topologicality
will require more detailed study.

We acknowledge comprehensive discussion with R. R. P. Singh.  This work was supported
by DOE Grant DE-FG02-04ER46111.


\end{document}